\documentclass[english,aps,manuscript]{revtex4}
\usepackage[T1]{fontenc}
\usepackage[latin9]{inputenc}
\usepackage{amsmath}
\usepackage{amssymb}
\usepackage{graphicx}
\usepackage{esint}

\makeatletter
\@ifundefined{textcolor}{}
{%
 \definecolor{BLACK}{gray}{0}
 \definecolor{WHITE}{gray}{1}
 \definecolor{RED}{rgb}{1,0,0}
 \definecolor{GREEN}{rgb}{0,1,0}
 \definecolor{BLUE}{rgb}{0,0,1}
 \definecolor{CYAN}{cmyk}{1,0,0,0}
 \definecolor{MAGENTA}{cmyk}{0,1,0,0}
 \definecolor{YELLOW}{cmyk}{0,0,1,0}
}

\usepackage{caption}

\@ifundefined{showcaptionsetup}{}{%
 \PassOptionsToPackage{caption=false}{subfig}}
\usepackage{subfig}
\makeatother

\usepackage{babel}
\begin{document}

\title{Free expansion of Bose-Einstein Condensates with a Multi-charged
Vortex}

\author{R. P. Teles, F. E. A. dos Santos, M. A. Caracanhas, and V. S. Bagnato}

\affiliation{Instituto de F�sica de S�o Carlos, USP, Caixa Postal 369, 13560-970
S�o Carlos, S�o Paulo, Brazil}
\begin{abstract}
In this work, we analyze the free expansion of Bose-Einstein condensates
containing multi-charged vortices. The atomic cloud is initially confined
in a three-dimensional asymmetric harmonic trap. We apply both approximate
variational solutions and numerical simulations of the Gross-Pitaevskii
equation. The data obtained provide a way to establish the presence
as well as the multiplicity of vortices based only on the properties
of the expanded cloud which can be obtained via time-of-flight measurements.
In addition, several features like the evolution of the vortex core
size and the asymptotic velocity during free expansion were studied
considering the atomic cloud as being released from different harmonic
trap configurations. 
\end{abstract}
\maketitle

\section{Introduction}

The extension of quantum phenomena into macroscopic scales is responsible
for a whole class of effects such as superconductivity, superfluidity,
and Bose-Einstein Condensation, which played central roles in the
last-century physics. The production of the first Bose-Einstein condensates
(BECs), using rubidium \cite{anderson} and sodium \cite{davis} atoms,
turned possible the realization of experiments involving macroscopic
quantum phenomena with unprecedented level of control of the external
parameters.

Vortices in BECs are topological defects characterized by a quantized
angular momentum. A conventional method for generation of such defects
consists in confining the condensed atomic cloud into a rotating trap.
It turns out that, for angular velocities higher than a critical value
$\Omega_{c}$, vortex states become energetically favorable, thus
inducing the creation of quantized vortices \cite{critical1,vf1,vor2,pd}.
Experimental realizations of condensed alkali atoms confined by more
general time-dependent potentials allowed the observation not only
of vortex lattices but also of quantum turbulence \cite{3v,rev3,turb,pethick,davis}.
Since quantum turbulence is characterized by the presence of a self-interacting
tangle of quantized vortices, the correct understanding of dynamics,
formation, and stability of vortices have shown to be of paramount
importance \cite{turb,liv3,d1v,vi1,vi2} being the subject of many
theoretical works \cite{aa,fad,obv,pethick,rev2,dsv}. In particular,
the role of acoustic excitations generated by decaying multi-charged
vortices in the development of turbulence is still an open question
\cite{PRB53-75}. This work aims to provide a set of tools that helps
to identify the presence as well as the charge of vortices in both
turbulent and non-turbulent clouds observed using time-of-flight pictures.

The radius of vortex cores are typically of the order of the healing
length of the condensate. Such a small size makes \emph{in situ} observations
very difficult. The most common method for visualization of vortices
in BECs relies on the so called time-flight pictures which can be
obtained after releasing the condensed cloud from its trap and letting
it expand freely for some time, typically tens of milliseconds \cite{vp,vor1,rev3,obv,MIT}.
To determine the charge multiplicity of vortices in confined clouds
using time-of-flight pictures, it is necessary to establish the correct
connection between the features of the trapped and expanded clouds.

Charged vortices have regions of stability which depend on some of
the sample characteristics such as: winding number, trap anisotropy
and intensity of the atomic interaction\cite{stab01,stab02,stab03,mcv03,mcv01,mcv06}.
The majority of the theories of multi-charged vortex stability do
not take into account the presence of thermal cloud. In realistic
experiments, the coupling between condensate and thermal cloud is
an important key to stability. In general, dissipation makes multi-charged
vortices split into singly charged vortices\cite{mcv07,mcv03}. However,
some research groups have implemented techniques to overcome this
issue, as for example: implementing a tightly focused resonant laser\cite{mcv02},
using a blue-detuned laser beam which compensates the gravity\cite{mcv04},
and by the application of a Gaussian potential peak along the vortex
core \cite{mcv05}.

In this paper, we considered the expansion of a BEC containing a multiply
charged vortex at its center. The main calculations are performed
by using a variational method which takes into account the presence
of non-fundamental vortices. A similar work for single charged vortices
in two-dimensional condensates was done by Lundh \emph{et al. }\cite{emil}
and a more numerical approach was employed by Dalfovo \emph{et al.}
in Ref. \cite{michele}.

This work is divided as follows: section I is this introduction. Section
II presents the variational method used in this work. In the section
III, we discuss the dynamical equations for the expanding condensate.
Finally, in section IV there is a general discussion of our results.

\section{Variational Method}

At zero temperature, a Bose gas with scattering length $a_{s}$ much
smaller than the average interparticle distance can be described by
the Gross-Pitaevskii equation (GPE)
\begin{equation}
i\hbar\frac{\partial\Psi(\mathbf{r},t)}{\partial t}=\left[-\frac{\hbar^{2}}{2m}\nabla^{2}+V(\mathbf{r})+U_{0}\left|\Psi(\mathbf{r},t)\right|^{2}\right]\Psi(\mathbf{r},t),\label{eq:1.1}
\end{equation}
with the harmonic trap in cylindrical coordinates given by $V(\mathbf{r})=\frac{1}{2}m\omega_{\rho}^{2}(\rho^{2}+\lambda^{2}z^{2})$,
where $m$ is the mass of the particles, $\lambda$ is an anisotropy
parameter, and the coupling constant is $U_{0}=4\pi\hbar^{2}a_{s}/m$.
Following the variational principle, the Lagrangian density $\mathcal{L}$
which recovers the GPE for a complex field $\Psi(\mathbf{r},t)$ can
be written as
\begin{equation}
\begin{array}{c}
\mathcal{L}={\displaystyle \frac{i\hbar}{2}\left(\Psi^{*}(\mathbf{r},t)\frac{\partial\Psi(\mathbf{r},t)}{\partial t}-\Psi(\mathbf{r},t)\frac{\partial\Psi^{*}(\mathbf{r},t)}{\partial t}\right)}\\
{\displaystyle -\frac{\hbar^{2}}{2m}\left|\nabla\Psi(\mathbf{r},t)\right|^{2}-V(\mathbf{r})\left|\Psi(\mathbf{r},t)\right|^{2}-\frac{U_{0}}{2}\left|\Psi(\mathbf{r},t)\right|^{4}}.
\end{array}\label{eq:1.2}
\end{equation}

In the variational method, the wave function $\Psi(\mathbf{r},t)$
of a condensate containing one central vortex with charge $\ell$
is approximated by a trial function $\Psi_{\ell}(\mathbf{r},t)$ which
depends on a set of variational parameters $q_{i}=q_{i}(t)$ \cite{perez1,perez2}.
This function can then be substituted into Lagrange function

\begin{equation}
L=\int\mathcal{L}d^{3}\mathbf{r}.\label{eq:1.3}
\end{equation}
This way, the time evolution of the parameters $q_{i}$ follows the
Euler-Lagrange equations

\begin{equation}
\frac{\partial}{\partial t}\left(\frac{\partial L}{\partial\dot{q}_{i}}\right)-\frac{\partial L}{\partial q_{i}}=0.\label{eq:1.4}
\end{equation}

Here, we generalize the Gaussian trial function in Ref. \cite{emil}
to the case of three-dimensional BECs with multiply charged vortices

\begin{equation}
\Psi_{\ell}(\rho,\phi,z,t)=\left[\frac{N}{\pi^{\frac{3}{2}}\ell!R_{\rho}(t)^{2\ell+2}R_{z}(t)}\right]^{\frac{1}{2}}\rho^{\ell}e^{i\ell\phi}\psi_{G}(\rho,z,t)e^{iB_{\rho}(t)\frac{\rho^{2}}{2}+iB_{z}(t)\frac{z^{2}}{2}},\label{eq:2.3}
\end{equation}
with the function $\psi_{G}$ being given by

\begin{equation}
\psi_{G}(\rho,z,t)=\exp\left(-\frac{\rho^{2}}{2R_{\rho}(t)^{2}}\right)\exp\left(-\frac{z^{2}}{2R_{z}(t)^{2}}\right).\label{eq:2.4}
\end{equation}

If we consider $\ell=0$ , we recover the vortex-free approximation
proposed by P�rez-Garc�a \emph{et al}. in Ref. \cite{perez1}. In
our case, however, the parameter $R_{\rho}$ is no longer the mean
square root of $\rho$. Instead, it is related to $\sqrt{\left\langle \rho^{2}\right\rangle }$
according to

\begin{equation}
\sqrt{\left\langle \rho^{2}\right\rangle }=\sqrt{\ell+1}R_{\rho}.
\end{equation}

Here we define the vortex core $\xi$ as the healing length calculated
at the center of the condensate without the central vortex. This assumption
leads us to

\begin{equation}
\xi=\ell\hbar\pi^{\frac{3}{4}}R_{\rho}\sqrt{\frac{R_{z}}{2mNU_{0}}}.\label{eq:2.6}
\end{equation}

\section{dynamical equations}

By substituting (\ref{eq:2.3}) into Eqs. (\ref{eq:1.2}) and (\ref{eq:1.3}),
and then performing the spacial integrations, we obtain the Lagrange
function for the variational parameters 
\begin{equation}
L=-N\hbar\omega_{\rho}\left\{ \frac{\left(\ell+1\right)}{2}\left[\frac{1}{r_{\rho}^{2}}+\left(\dot{\beta}_{\rho}+\beta_{\rho}^{2}+1\right)r_{\rho}^{2}\right]+\frac{1}{4}\left[\frac{1}{r_{z}^{2}}+\left(\dot{\beta}_{z}+\beta_{z}^{2}+\lambda^{2}\right)r_{z}^{2}\right]+\frac{\gamma(2\ell)!}{2^{2\ell}\sqrt{2\pi}\left(\ell!\right)^{2}r_{\rho}^{2}r_{z}}\right\} ,\label{eq:2.8}
\end{equation}
where the parameters were rescaled according to $R_{\rho}(t)=a_{osc}r_{\rho}(t)$,
$R_{z}(t)=a_{osc}r_{z}(t)$, $B_{\rho}(t)=a_{osc}^{-2}\beta_{\rho}(t)$,
and $B_{z}(t)=a_{osc}^{-2}\beta_{z}(t)$. The harmonic oscillator
length was defined as $a_{osc}=\sqrt{\hbar/m\omega_{\rho}}$, whereas
the dimensionless iteration was defined according to $\gamma=Na_{s}/a_{osc}$.
The Euler-Lagrange equations \eqref{eq:1.4} applied to the Langrangean
\eqref{eq:2.8} finally give us the equations of motion
\begin{align}
\left(\dot{\beta}_{\rho}+\beta_{\rho}^{2}+1\right)r_{\rho}= & \frac{1}{r_{\rho}^{3}}+\frac{\gamma(2\ell)!}{2^{2\ell-1}\sqrt{2\pi}\left(\ell+1\right)!\ell!r_{\rho}^{3}r_{z}},\label{eq:2.9a}\\
\beta_{\rho}= & {\displaystyle \frac{\dot{r_{\rho}}}{r_{\rho}}},\label{eq:2.9b}\\
\left(\dot{\beta}_{z}+\beta_{z}^{2}+\lambda^{2}\right)r_{z}= & \frac{1}{r_{z}^{3}}+\frac{\gamma(2\ell)!}{2^{2\ell-1}\sqrt{2\pi}\left(\ell!\right)^{2}r_{\rho}^{2}r_{z}^{2}},\label{eq:2.9c}\\
\beta_{z}= & {\displaystyle \frac{\dot{r_{z}}}{r_{z}}}.\label{eq:2.9d}
\end{align}
By taking the time-derivative of Eqs. (\ref{eq:2.9b}) and (\ref{eq:2.9d}),
the parameters $\beta_{\rho}$ and $\beta_{z}$ can be eliminated
in such a way that these four equations can be reduced to the following
two:
\begin{align}
\ddot{r_{\rho}}+r_{\rho}= & \frac{1}{r_{\rho}^{3}}+\frac{\gamma(2\ell)!}{2^{2\ell-1}\sqrt{2\pi}\left(\ell+1\right)!\ell!r_{\rho}^{3}r_{z}},\label{eq:2.10a}\\
\ddot{r_{z}}+\lambda^{2}r_{z}= & \frac{1}{r_{z}^{3}}+\frac{\gamma(2\ell)!}{2^{2\ell-1}\sqrt{2\pi}\left(\ell!\right)^{2}r_{\rho}^{2}r_{z}^{2}}.\label{eq:2.10b}
\end{align}

\subsection{Free expansion}

By considering the stationary solution for the Eqs. of motion (\ref{eq:2.10a})
and (\ref{eq:2.10b}), we obtain the algebraic equations

\begin{align}
r_{\rho0}^{4}= & 1+\frac{\gamma(2\ell)!}{2^{2\ell-1}\sqrt{2\pi}(\ell+1)!\ell!r_{z0}}\label{eq:3.2a}\\
\lambda^{2}r_{z0}^{4}= & 1+\frac{\gamma(2\ell)!r_{z0}}{2^{2\ell-1}\sqrt{2\pi}(\ell!)^{2}r_{\rho0}^{2}}.\label{eq:3.2b}
\end{align}

The free expansion equations are obtained when the trap terms in Eqs.
\eqref{eq:2.10a} and \eqref{eq:2.10b} are neglected, and thus they
become

\begin{align}
\ddot{r_{\rho}}= & \frac{1}{r_{\rho}^{3}}+\frac{(2\ell)!\gamma}{2^{2\ell-1}\left(2\pi\right)^{\frac{1}{2}}(\ell+1)!\ell!r_{\rho}^{3}r_{z}},\label{eq:4.2a}\\
\ddot{r_{z}}= & \frac{1}{r_{z}^{3}}+\frac{(2\ell)!\gamma}{2^{2\ell-1}\left(2\pi\right)^{\frac{1}{2}}(\ell!)^{2}r_{\rho}^{2}r_{z}^{2}}.\label{eq:4.2b}
\end{align}

The first and second terms in the r.h.s of (\ref{eq:4.2a}) and (\ref{eq:4.2b})
come from the kinetic and interaction terms in (\ref{eq:1.2}), respectively.
From (\ref{eq:4.2a}) and (\ref{eq:4.2b}), we can also observe that
the interaction term is dominant in the long-time limit, while the
kinetic terms plays a role only at the first milliseconds of the expansion.
This is however not the case when we consider extremely large values
of $\ell$. In this case, the interaction terms can be neglected thus
leading to equations identical to the ones describing the free expansion
of an ideal gas
\begin{align}
\ddot{r_{\rho}}= & \frac{1}{r_{\rho}^{3}},\label{eq:4.3a}\\
\ddot{r_{z}}= & \frac{1}{r_{z}^{3}},\label{eq:4.3b}
\end{align}
which have the simple solution
\begin{equation}
r_{j}(\tau)=\sqrt{r_{j}(0)^{2}+r_{j}(0)^{-2}\tau^{2}},\label{eq:4.4}
\end{equation}
with $j=\rho,z$.

\begin{figure}
\centering
\subfloat[Prolate shape ($\lambda=0.1$).]{\includegraphics[scale=0.65]{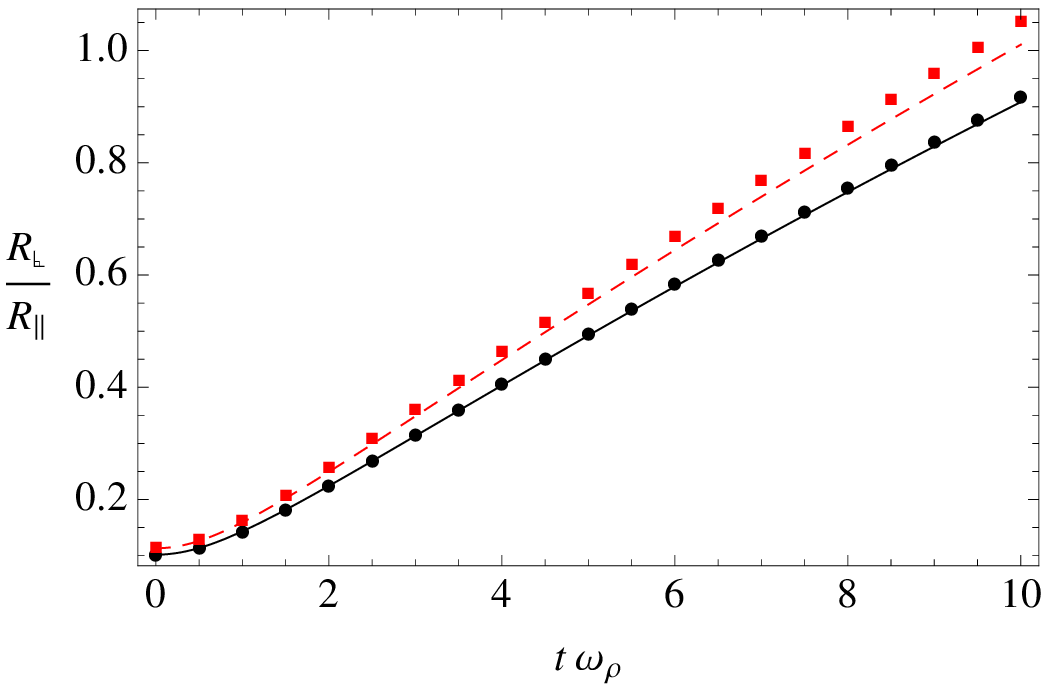}

}\subfloat[Spherical shape ($\lambda=1$).]{\includegraphics[scale=0.65]{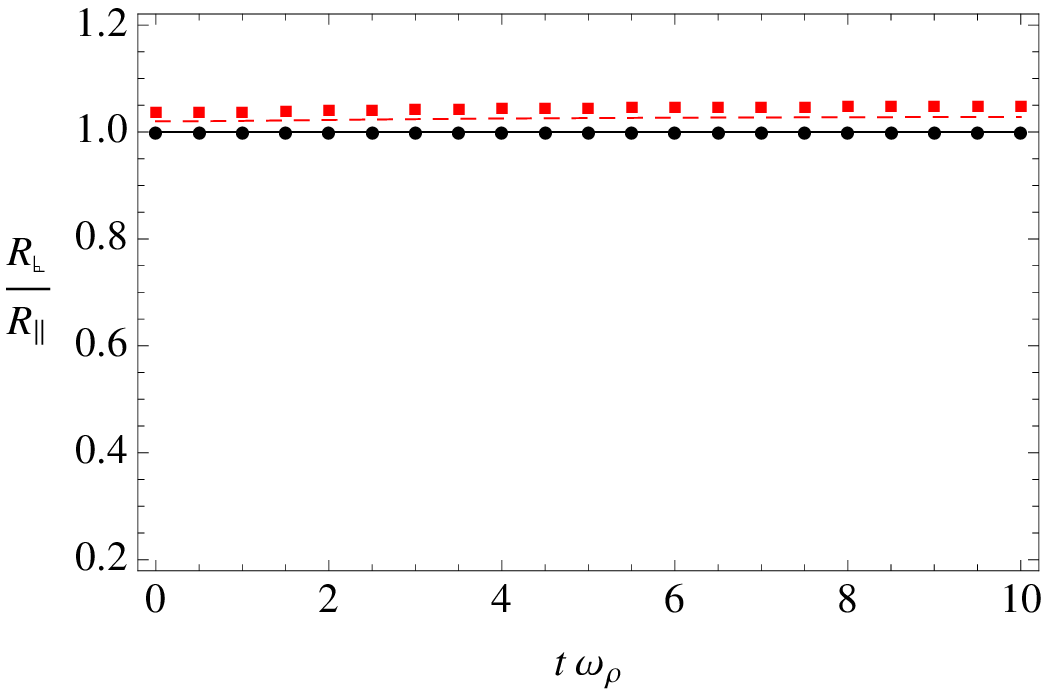}

}

\caption{(Color online) Comparison of the free expansion for $\gamma=800$
using the variational method and the the direct numerical simulation
of GPE. The lines and marks come from the variational method and the
numerical simulation of GPE, respectively. Full line and circle mark
(black color) correspond to $\ell=0$, and the dashed line and square
mark (red color) correspond to $\ell=2$.}

\label{fig:2}
\end{figure}

In order to check the accuracy of our results, we performed direct
simulation of the GP equation using the Fourier spectral method in
space where the Fourier components of $\Psi(\mathbf{r},t)$ where
computed using fast Fourier transformations\cite{xmds}. In figure
\ref{fig:2}, our variational method is compared with the high-precision
numerical simulation for expanding spherical as well as prolate condensates. 

\section{Results and Discussions}

\begin{figure}
\centering

\includegraphics{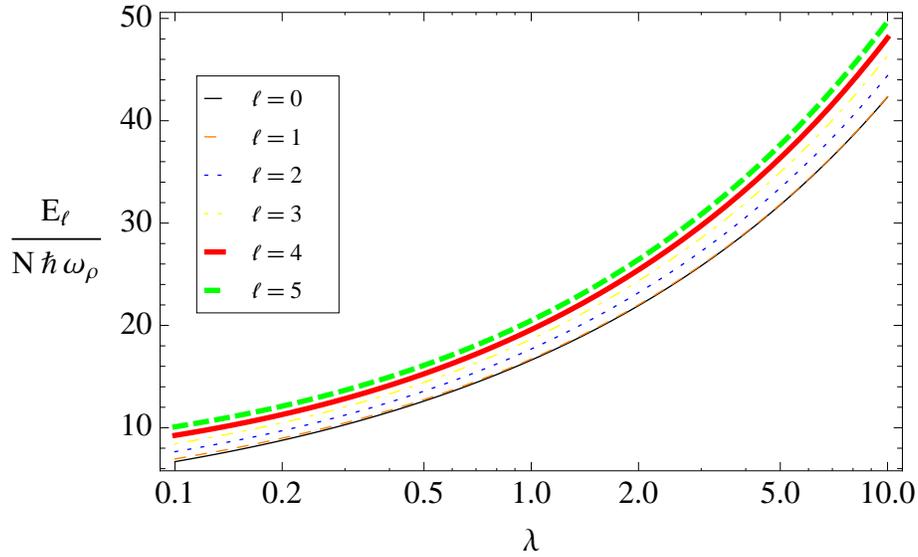}

\caption{(Color online) Energy plotted as function of the trap anisotropy for
different vortex circulations. The dimensionless interaction parameter
is taken as $\gamma=800$. This corresponds to a condensate with $10^{5}$
atoms and s-wave scattering length $a_{s}=100a_{0}$.}

\label{fig:3}
\end{figure}

\begin{figure}
\centering
\captionsetup{labelfont=bf}

\subfloat[$\lambda=0.1$]{\includegraphics[scale=0.65]{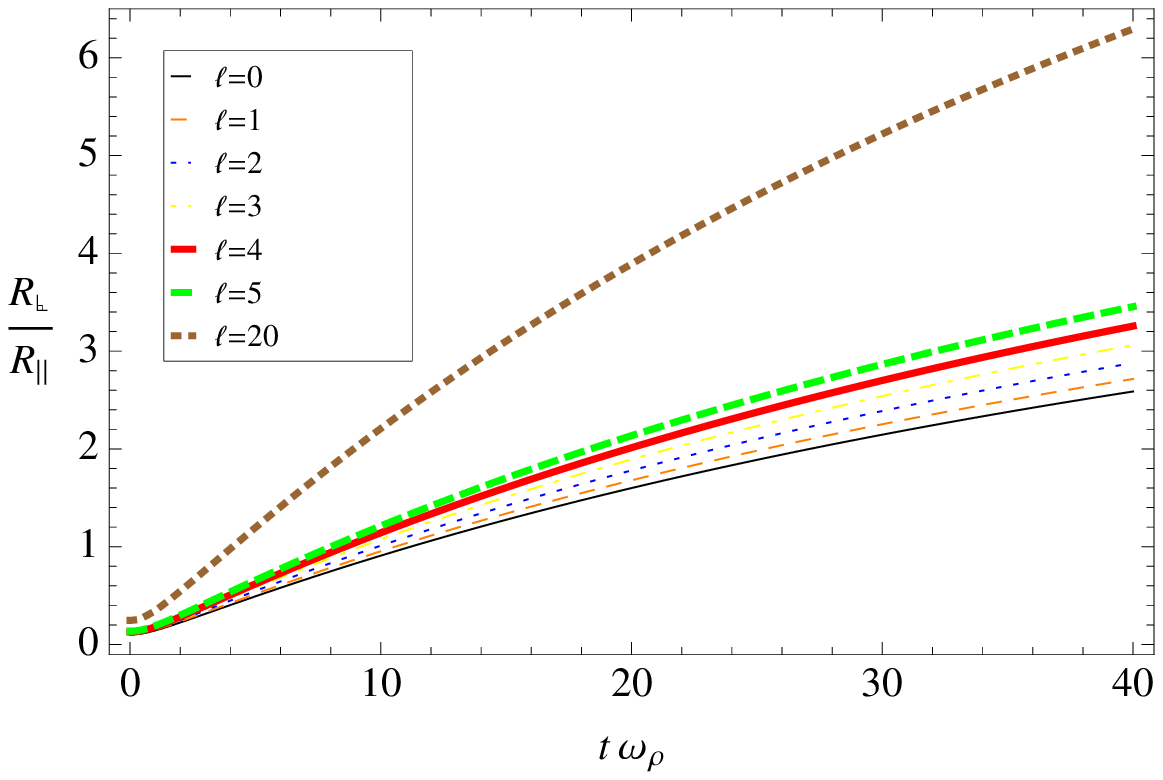}

\label{fig:4-a}}\subfloat[$\lambda=10$]{\includegraphics[scale=0.65]{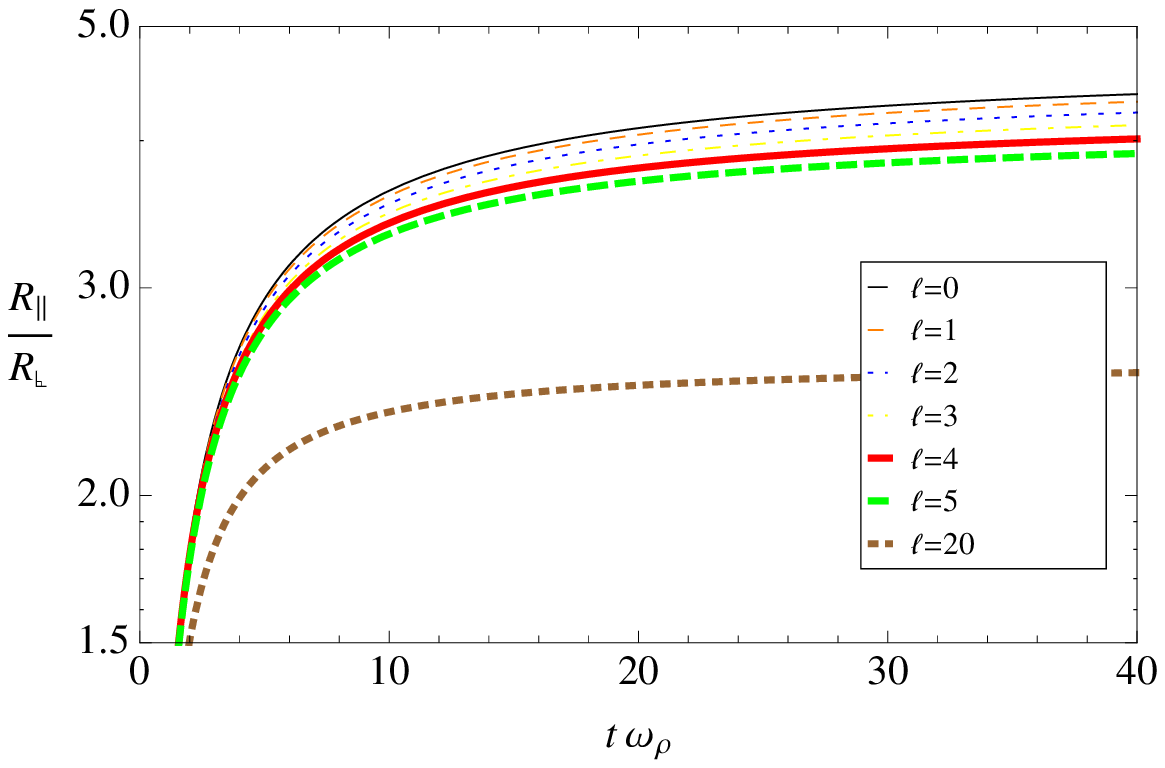}

\label{fig:4-b}}

\subfloat[$\lambda=1$]{\includegraphics[scale=0.65]{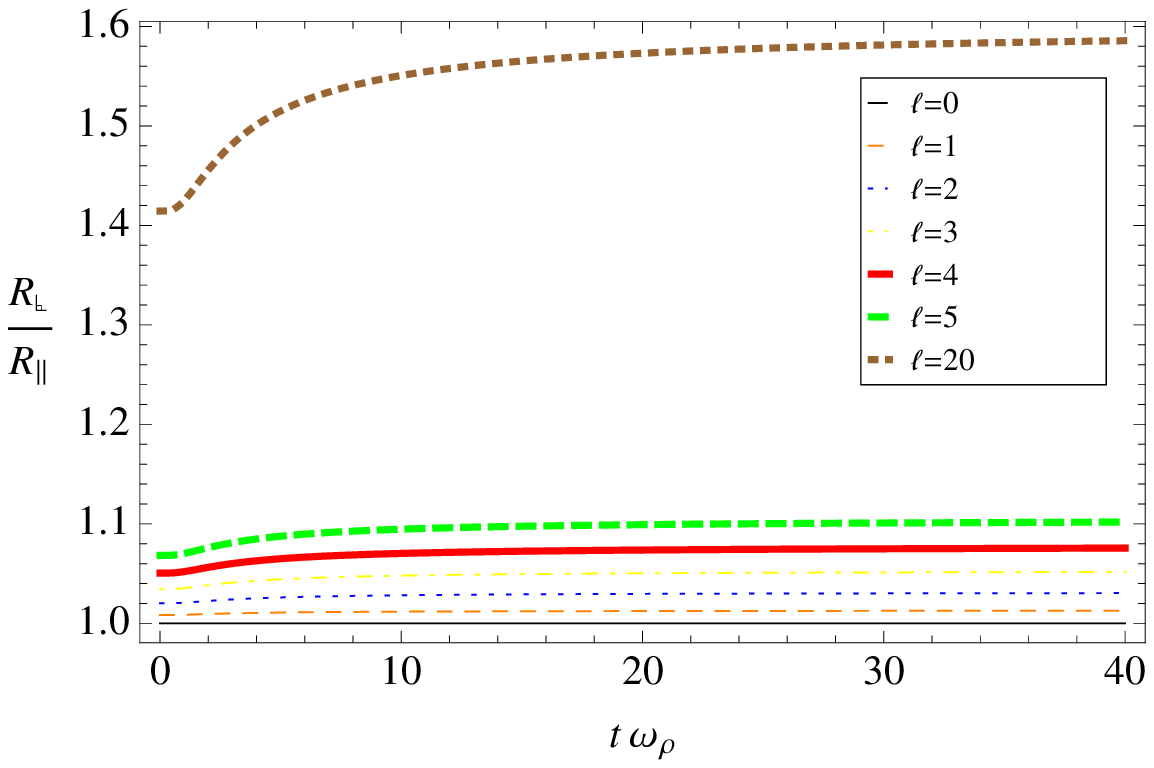}

\label{fig:4-c}}\subfloat[$\ell=1$]{\includegraphics[scale=0.65]{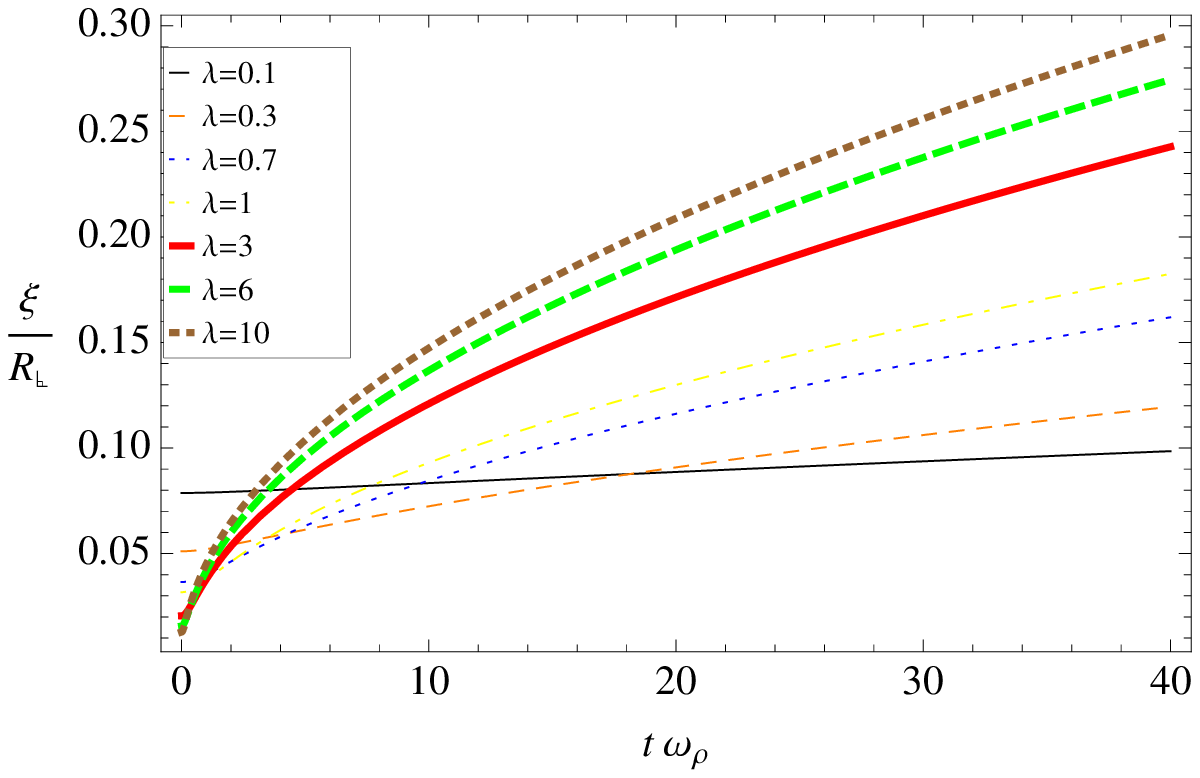}

\label{fig:4-d}}

\caption{(Color online) Aspect ratio evolution during free expansion for the
respective trapped shapes: (a) prolate, (b) oblate and (c) spherical.
They were calculated for $\gamma=800$. Frame (d) represents the ratio
between the vortex core and the radial radius during the expansion
for several kind of trap shapes.}

\label{fig:4}
\end{figure}

\begin{figure}
\subfloat[$\lambda=0.1$]{\includegraphics[scale=0.65]{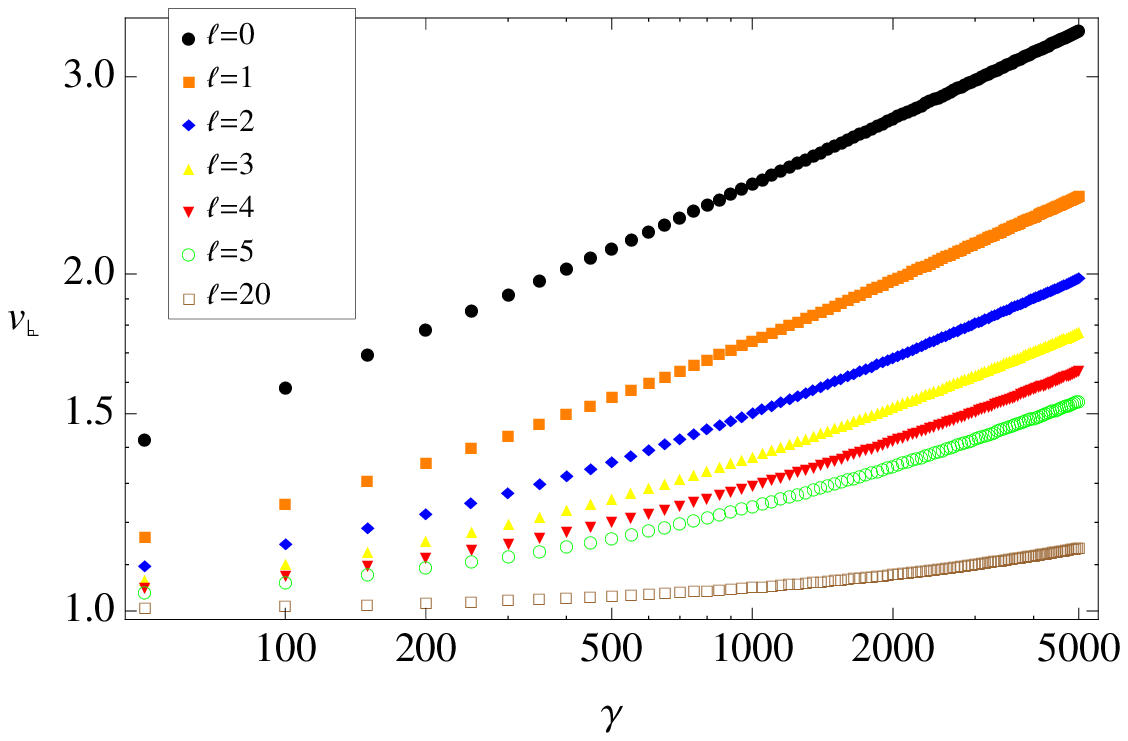}\includegraphics[scale=0.65]{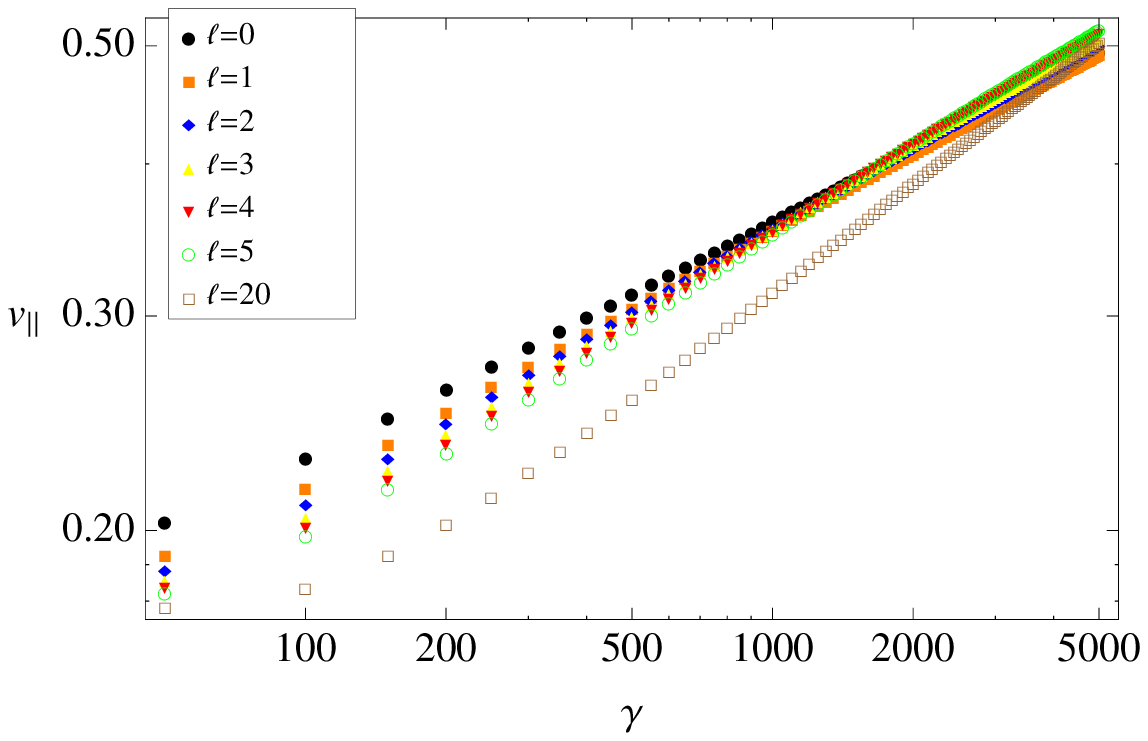}

}

\subfloat[$\lambda=10$]{\includegraphics[scale=0.65]{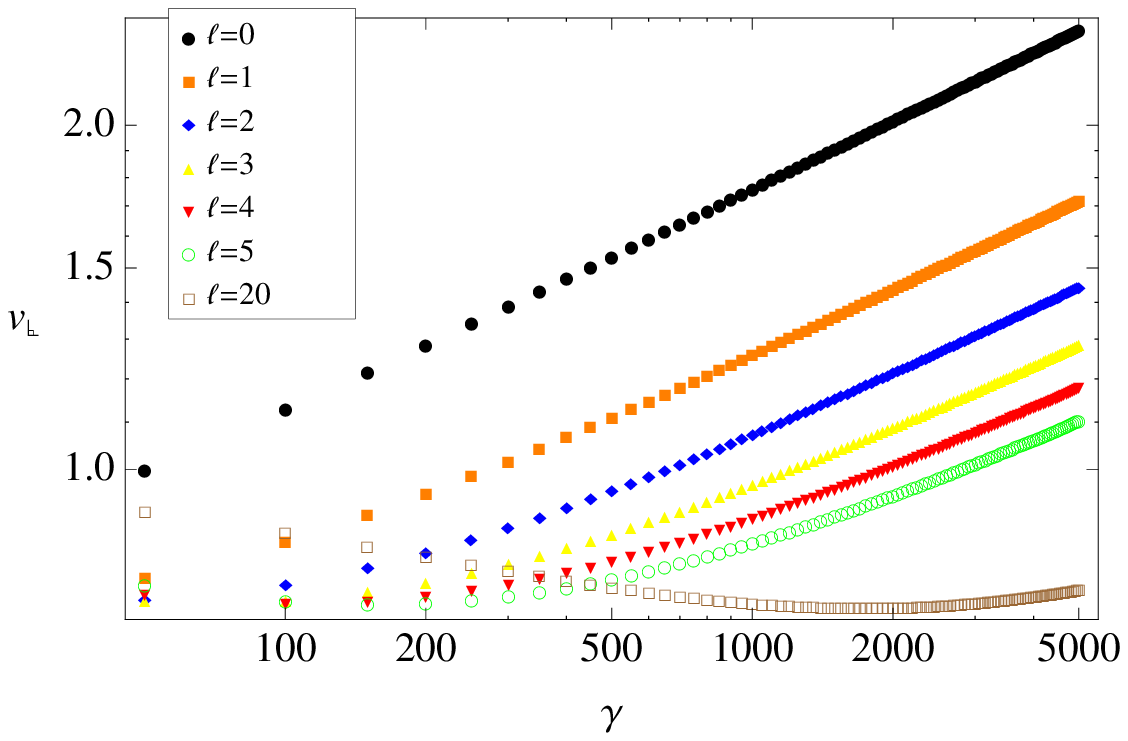}\includegraphics[scale=0.65]{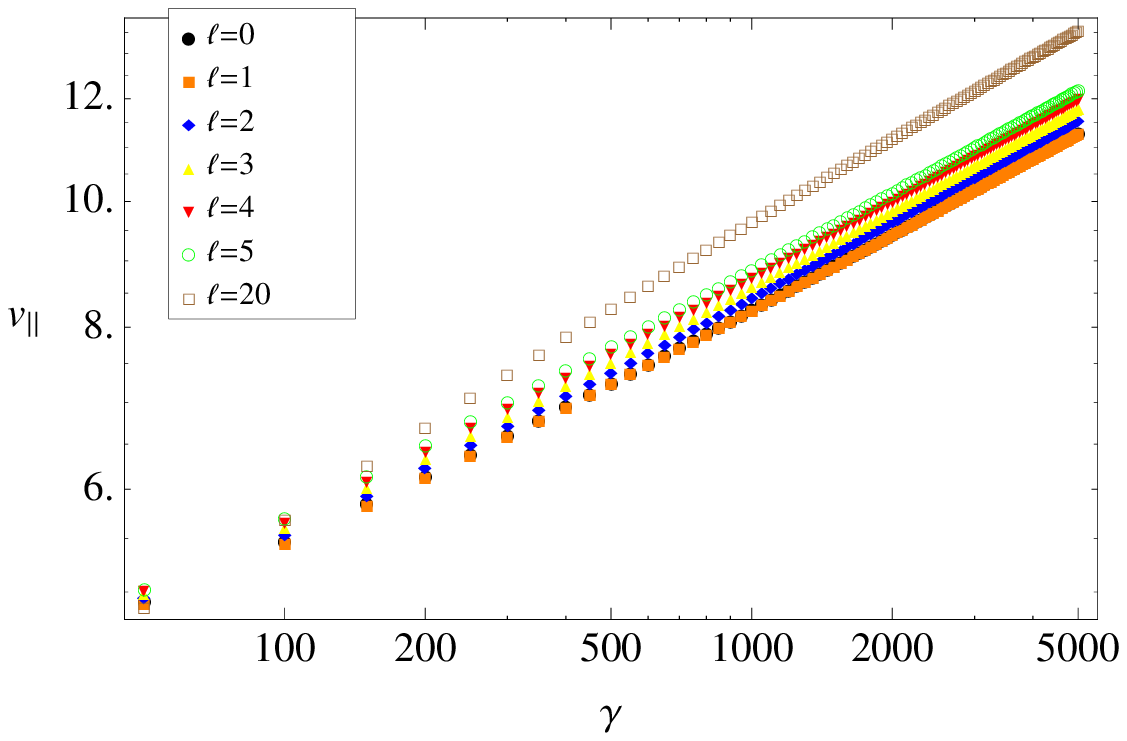}

}

\caption{(Color online) The asymptotic behavior of expansion velocities as
a function of $\gamma$ for initially prolate (a) and oblate (b) shapes.}

\label{fig:5}
\end{figure}

The initial condition for the variational parameters where calculated
from Eqs. \eqref{eq:3.2a} and \eqref{eq:3.2b} considering the radial
frequency $\omega_{\rho}=2\pi\times207Hz$. The time evolution of
the parameters was obtained by numerically solving Eqs. \eqref{eq:4.2a}
and \eqref{eq:4.2b} using the fourth-order adaptive Runge-Kutta method,
with expansion times of the order of $30ms$. These values where chosen
in order to be consistent with our experiments using $^{87}$Rb 87
atoms \cite{emanuel}.

Figure \ref{fig:3} shows the initial energy of the system for a fixed
interaction parameter as function of the trap anisotropy measured
by the parameter $\lambda$. It shows the monotonic growth of the
condensate energy with the circulation of the vortex due to the extra
kinetic energy and the larger volume occupied by the condensate with
increasing $\ell$.

The vortex core evolution was analyzed considering its size to be
given by Eq. \eqref{eq:2.6}. A disadvantage of this method comes
from the fact that the static constraint \eqref{eq:2.6} between the
vortex core and the cloud dimensions is considered to be also valid
during the entire expansion time. At least in principle, our Ansatz
can be improved by introducing an additional parameter which characterizes
the core expansion independently. In practice, the introduction of
such an additional parameter affecting the density profile of the
cloud is a not a trivial task. Indeed, the phase of the condensate
wave function must also be modified in order to reproduce superfluid
current corresponding to the time-variant core size. However, since
in this work our attention is restricted to static configurations
and the expansion of the cloud, the approach used here work turns
out to be appropriate as we can see from Fig. \ref{fig:2}.

In Fig. \ref{fig:4}(a)--(c), the time evolution of the aspect ratio
for different trap configurations and vortex circulations is depicted.
Figure \ref{fig:4}(c) shows how the growth of the dimensions of the
cloud during the expansion are affected by the circulation of the
central vortex. The higher the circulation of these vortices, the
greater is the anisotropy introduced in the cloud shape. This effect
is further increased during the cloud expansion.

In prolate condensates, Fig. \ref{fig:4-a} shows that the circulation
has the effect of decreasing the time for the aspect ratio inversion
due to the larger velocity field in the plane perpendicular to the
vortex line before the expansion. In the opposite case, where the
initial geometry is oblate, as in Fig. \ref{fig:4-b}, the circulation
increases the time required for aspect ratio inversion.

In this work, the expansion of the vortex core is analyzed by evolving
both the radial and axial radius using Eqs. \eqref{eq:4.2a} and \eqref{eq:4.2b},
and then calculating the radius $\xi$ of the vortex according to
\eqref{eq:2.6}. By taking the asymptotic solutions of Eqs. \eqref{eq:4.2a}
and \eqref{eq:4.2b}, it was also possible to extract the asymptotic
expansion velocities along the radial and axial directions, as shown
in Fig. \ref{fig:5}. These information supply a method for determining
the presence as well as the circulation of central vortices in an
asymmetric cloud by simply looking at the asymptotic expansion velocities,
which would require the repetition of the experiment considering different
expansion time. An alternative method relies on dependence of the
expansion dynamics on the circulation $\ell$ as depicted in Fig.
\ref{fig:4-d}. In fact, Figs. \ref{fig:4}(a)--(c) give a one to
one correspondence between the aspect ratio of the expanded cloud
and the multiplicity of the central vortex. This way, the multiplicity
of the central core can be determined from aspect ratio obtained from
time-of-flight pictures.
\begin{acknowledgments}
We acknowledge the financial support of from the National Council
for the Improvement of Higher Education (CAPES) and from the State
of S\~ao Paulo Foundation for Research Support (FAPESP).
\end{acknowledgments}
\pagebreak{}

\bibliographystyle{apsrev}
\bibliography{refdoc.bib}

\end{document}